\documentclass[journal]{vgtc}                

\ifpdf
  \pdfoutput=1\relax                   
  \usepackage{graphicx}                
  \DeclareGraphicsExtensions{.pdf,.png,.jpg,.jpeg} 
\else
  \usepackage{graphicx}                
  \DeclareGraphicsExtensions{.eps}     
\fi%

\graphicspath{{figures/}{pictures/}{images/}{./}} 

\usepackage{microtype}                 
\PassOptionsToPackage{warn}{textcomp}  
\usepackage{textcomp}                  
\usepackage{mathptmx}                  
\usepackage{times}                     
\usepackage{cite}                      
\usepackage{tabu}                      
\usepackage{booktabs}                  

\usepackage{color}
\usepackage{booktabs}
\usepackage{textcomp}
\usepackage{enumerate}
\usepackage{multirow}
\usepackage[dvipsnames]{xcolor}
\usepackage{subcaption}
\usepackage{hhline}
\usepackage[normalem]{ulem}

\newenvironment{myquote}%
  {\list{}{\leftmargin=0.2in\rightmargin=0.2in}\item[]}%
  {\endlist}

\onlineid{0}

\vgtccategory{Research}
\vgtcpapertype{please specify}

\title{Touching Data: A Discoverability-based Evaluation of a Visualization Interface for Tablet Computers}

\author{Ramik Sadana, Meeshu Agnihotri, and John Stasko}
\authorfooter{
\item
 Ramik Sadana, Meeshu Agnihotri and John Stasko are with Georgia Tech. E-mail: \{ramik,meeshu\}@gatech.edu, stasko@cc.gatech.edu.
}

\shortauthortitle{Biv \MakeLowercase{\textit{et al.}}: Global Illumination for Fun and Profit}

\abstract{While a number of touch-based visualization systems have appeared in recent years, relatively little work has been done to evaluate these systems. The prevailing methods compare these systems to desktop-class applications or utilize traditional training-based usability studies. We argue that existing studies, while useful, fail to address a key aspect of mobile application usage -- initial impression and discoverability-driven usability. Over the past few years, we have developed a tablet-based visualization system, Tangere, for analyzing tabular data in a multiple coordinated view configuration. This article describes a discoverability-based user study of Tangere in which the system is compared to a commercially available visualization system for tablets -- Tableau's Vizable. The study highlights aspects of each system's design that resonate with study participants, and we reflect upon those findings to identify design principles for future tablet-based data visualization systems.%
} 

\keywords{Evaluation, tablet computers, touch interaction, data visualization, discoverability, memorability}

\CCScatlist{ 
 \CCScat{K.6.1}{Management of Computing and Information Systems}%
{Project and People Management}{Life Cycle};
 \CCScat{K.7.m}{The Computing Profession}{Miscellaneous}{Ethics}
}

\renewcommand{\manuscriptnotetxt}{}

\begin{document}


\firstsection{Introduction}

\maketitle


With the growing popularity and use of portable multi-touch devices such as tablet computers, several commercial~\cite{noauthor_roambi_2014,noauthor_vizable_2015,noauthor_business_nodate, noauthor_jmp_nodate} and research~\cite{baur_touchwave:_2012, jeffrey_m._rzeszotarski_kinetica:_2014, jo_touchpivot:_2017} applications providing information visualization capabilities on these devices have appeared. 
This development is still in its nascency, however. Most systems in this area look and feel very different from each other, adopt considerably varied interaction styles, and follow different design principles. Thus, fundamental questions about how touch-based interfaces affect visualization, and ultimately data comprehension, remain to be answered. 

During the past several years, we have designed a tablet-based visualization system, Tangere, for analyzing tabular data using diverse visualization techniques in a multiple coordinated view configuration~\cite{sadana_designing_2014,sadana_designing_2016}. The system provides operations for selecting items, zooming and panning, filtering data, examining data attributes, among others. To create the system, we identified the needs that tablet-based systems must address, defined requirements for the system, and designed, tested, and implemented the system to address those requirements. 


A logical next step is to evaluate the system. As we quickly learned, designing a study to robustly and sufficiently evaluate a tablet visualization system is a challenging task. Past research has favored methods that compare people's use of touch-based visualization applications to ones that are desktop-based ~\cite{jeffrey_m._rzeszotarski_kinetica:_2014,jo_touchpivot:_2017}. While informative, such studies exhibit a number of limitations and problems. First, desktop systems tend to be much more feature rich than tablet systems. A comparison thus requires restricting and constraining participants' usage of the desktop systems to only a subset of features they provide. 
Second, these studies tend to replicate results from the general HCI literature on differences between touch-based and cursor-based interfaces~\cite{hinckley_input_2002,wigdor_brave_2011}. Those studies found that experience with touch is considerably different than with cursor, participants prefer touch input over cursor input, and each method is better suited for specific types of tasks. 

An alternative evaluation methodology could compare the Tangere system to another similar tablet-based visualization system and conduct a traditional usability study. In such a study, participants would be trained on each system and then presented with a series of analytical tasks to complete on some data set. We would measure task success and completion times, and log the different types of errors that were made. 

While a totally reasonable way of evaluating the system, we chose not to use this method. Clearly, it relies on participants receiving appropriate training on the interfaces. We were concerned, however, that in real world scenarios and use, such training likely does not occur and thus this type of study would suffer from low ecological validity.  Mobile device-based applications typically are distributed through app stores. We suspect that a majority of people who would use a system like Tangere would be unlikely to invoke training of the nature typically provided in a traditional usability study. A potential user's focus, instead, would be on accomplishing their real work tasks and they only would resort to tutorials and manuals as ``just-in-time'', last resort strategies~\cite{cockburn_supporting_2015,novick_why_2006,rieman_field_1996}. Familiarity in the wild, conversely, often occurs through trial and error over several sessions of extended usage. Such ``training'' is difficult to emulate in a controlled study.


Additionally, a key part of designing Tangere was optimizing it for unfamiliar users who are likely to approach the application without thorough training~\cite{sadana_designing_2014,sadana_designing_2016}. To assess Tangere as an effective tool for such scenarios, we
 chose to evaluate it via a discoverability-based evaluation methodology that examines the difficulty associated with identifying how to use a system and captures a person's initial impressions of the system. In this article, we explain this methodology and we describe an evaluation of Tangere and the commercial system Vizable using it. Our contributions are three-fold. 
First, we employ this discoverability-based evaluation methodology, based in part on the notion of ``flow'' in the context of initial usage, for visualization applications on tablets. Second, we present the study results that identify the strengths and weaknesses of the two systems and explore how study participants performed using each. Finally, we reflect on findings from the study, highlighting effective design decisions, in order to inform the creation of future systems in this space.  

\section{Related Work}
The visualization community has shown increasing interest in multitouch platforms recently.
Several commercial applications have appeared, either as companion systems for desktop-applications~\cite{noauthor_roambi_2014,noauthor_business_nodate, noauthor_jmp_nodate} or designed specifically for touch and tablets~\cite{noauthor_vizable_2015}. Interest has been equally promising within the research community. A growing number of applications target unique configurations (e.g., large displays~\cite{browne_data_2011,heilig_scattertouch:_2010,isenberg_collaborative_2009} or small handhelds~\cite{buering_user_2006}), specific visualization techniques (e.g., streamgraph~\cite{baur_touchwave:_2012}, scatterplot~\cite{jeffrey_m._rzeszotarski_kinetica:_2014,sadana_designing_2014}, barchart~\cite{drucker_touchviz:_2013}, and multiple coordinated views~\cite{sadana_designing_2016}), or diverse interaction styles~\cite{jo_touchpivot:_2017,sadana_expanding_2016}.

However, relatively little research on evaluating information visualization systems for touch-based, mobile platforms has been conducted. Isenberg et al.~\cite{isenberg_data_2013} review issues involved in creating visualizations for touch-based surfaces including the technical, social, and design challenges present. The authors call out evaluation as a specific challenge and particularly identify the challenge of teasing out an interactive surface's role in the efficiency and evaluation assessment.


In the past, researchers have employed a diverse range of techniques for evaluating  touch-based visualization applications. Baur et al.~\cite{baur_touchwave:_2012} present case studies to highlight the utility of their TouchWave technique. To explore effective interaction strategies for barchart visualizations on tablets, Drucker et al.~\cite{drucker_touchviz:_2013} design two  interfaces (WIMP-based and gesture-based) and compare participants' performance on the interfaces on low-level visualization tasks.
For Kinetica~\cite{jeffrey_m._rzeszotarski_kinetica:_2014}, the authors compare people' performance on the system to that on Microsoft Excel running on a desktop for basic data exploration and sensemaking tasks. 

More recently, Jo et al.~\cite{jo_touchpivot:_2017} employ a similar technique for TouchPivot, a tablet-based visualization system that uses pen-and-touch interactions to facilitate data comprehension on an interface containing both a data table and a visualization. They compare TouchPivot with two commercial desktop-based interfaces, Tableau and Microsoft Excel's PivotTable. Study results show that novices were both faster with TouchPivot and created a larger number of meaningful charts with it. While the above studies address handheld touch devices, explorations of visualization systems on large displays have used similar evaluation methodologies. Isenberg et al.~\cite{isenberg_exploratory_2010} and Lee at al.~\cite{lee_sketchinsight:_2015} evaluated participants' performance on predefined tasks and extracted both qualitative and quantitative metrics from these observations. Walny et al.~\cite{walny_understanding_2012} followed a similar approach, employing a Wizard of Oz methodology to simulate system response.  

These evaluation methods either employ performance-comparison techniques or use methods highlighting differences in experience between touch-based and cursor-based input. While they certainly generate insights on how touch alters participants' comprehension of data and usage patterns with visualization tools, they do not generate comprehensive insights pertaining to the design of systems in this space, e.g., best practices to amplify the effectiveness, discoverability and learnability of interactions. 

Discoverability is a well-accepted component of and contributor to the usability of an interface~\cite{norman_gestural_2010}. In the past, discoverability has also been described as ``initial learning.'' Nielsen defines this as a novice user's experience of the initial part of the learning curve~\cite{nielsen_usability_1994}, insisting that a highly learnable system could be categorized as ``allowing users to reach a reasonable level of usage proficiency within a short time.'' Shneiderman et al.~\cite{shneiderman_designing_2016} define it similarly as ``the time it takes members of the user community to learn how to use the commands relevant to a set of tasks.'' Others specifically identify the no-training aspect of initial learning with an interface. Butler~\cite{butler_connecting_1985} considers initial learnability as ``initial user performance based on self-instruction'', whereas Rieman~\cite{rieman_field_1996} assesses whether a software is ``minimally useful with no formal training.'' We use these concepts to examine the Tangere system to understand at a nuanced level the aspects of design that either amplify or diminish the usability of the system.

Appert et al.~\cite{appert_dwell-and-spring:_2012} used discoverability as an instrument for evaluating an interaction. The authors used it to assess the usability of a direct-manipulation gesture to perform undo and repo operations. We use a similar approach -- however, we magnify the scope from evaluating just a single interaction to evaluating an ensemble of interactions. 

A	 different study methodology that captures similar aspects of people's responses to an interface is the guessability study approach~\cite{wobbrock_user-defined_2009,kane_usable_2011,hinrichs_gestures_2011}. In this approach, participants are presented with an effect to be achieved and are asked to perform the interaction that they think would be appropriate to achieve this effect. The guessability study approach differs from discoverability in that the former is a formative evaluation while the latter is summative. However, both measure similar characteristics of people's familiarity and learned behavior, and help define (guessability) or assess (discoverability) the design of an interface. We did not employ guessability as the method cannot be used for comparing two interfaces.

\begin{figure*}[t!]
  \centering
  \includegraphics[width=2\columnwidth]{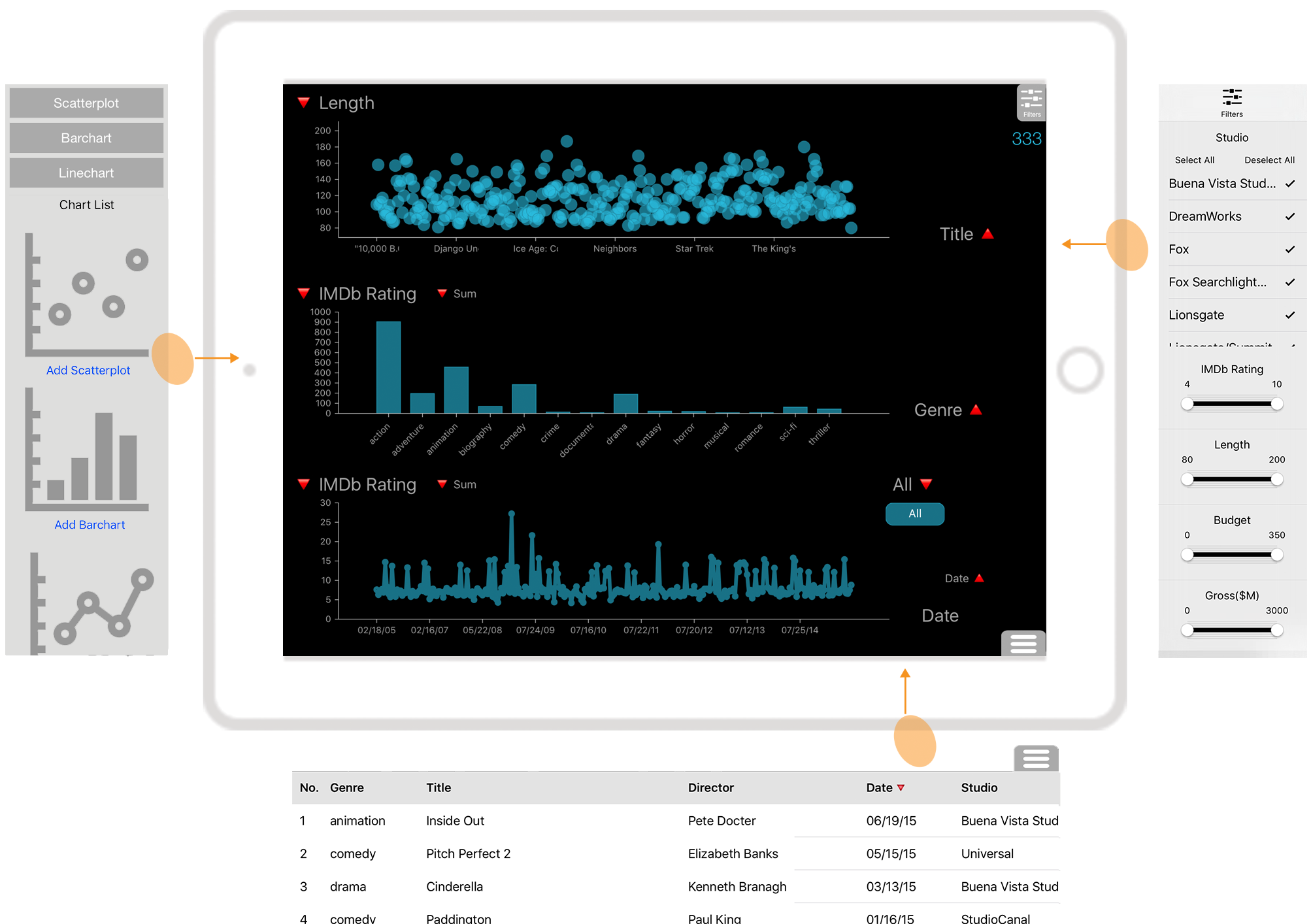}
  \caption{The Tangere system interface. The panels are placed beyond the screen and can be brought in with a gesture.}
  ~\label{fig:interface}
  \vspace{-1.5em}
\end{figure*}

The Tangere system~\cite{sadana_designing_2014,sadana_designing_2016}, shown in Figure~\ref{fig:interface}, runs on an Apple iPad tablet and is intended to help analyze tabular datasets. Data can be visualized using multiple techniques, including scatterplot, barchart, and linechart. Each technique supports a wide range of interactive operations such as selection, zooming, filtering, sorting (barcharts), and viewing details-on-demand, among others. Views also can appear in a coordinated configuration and support brushing \& linking. The primary goal of this research was to design a system that, while operating under the constraints of being touch-optimized, was still usable and effective as a standalone tool for data analysis. In this paper, we describe the evaluation techniques we used to assess how successfully the system achieves this goal.

\section{Designing an Evaluation for Tangere}
The overarching goal of this research was to assess a person's first experience using Tangere to understand its effectiveness as a tool for accomplishing traditional data analysis. More specifically, we sought to explore, understand, and highlight aspects of the system's design that either amplified or hampered usability. As described in~\cite{sadana_designing_2014,sadana_designing_2016}, the design of Tangere emerged through a process of prototyping and testing different options for each interaction. These options were influenced by a variety of sources, including existing applications, previous research, formative studies, and designer-generated options. 
Users of the system derive value out of how these options work in conjunction with each other to assist their workflow. Hinckley et al.~\cite{hinckley_pen_2010} term this the \textit{flow} of interaction and highlight the importance of evaluating it across a series of interrelated subtasks. We identified the analysis of this flow as the method for measuring Tangere's effectiveness.

Several metrics, both quantitative and qualitative, can be used to assess the flow. These include perceived effectiveness, efficiency, discoverability, learnability and memorability, among others. 
To capture these metrics, people need to engage the system and we must observe their behavior as they adapt to the experience and react to its idiosyncrasies. 

One reason a qualitative evaluation of the design is useful today is the infancy of the space of touch-based visualization systems. Researchers are still exploring potential solutions, and a breadth of choices exist for the types of systems that can be designed. In this work, we were less interested in identifying a ``type'' of system that can work best, as there may not be one specific type of system that meets all needs. 
Instead, our goal was to identify effective patterns and practices within systems being built in this space that could be adapted into future systems' designs. 
Specifically, we wanted to capture patterns that surface answers to broader questions, including how a design encourages gaining expertise and familiarity, and how it assists people to translate knowledge from other applications and platforms. 

Answers to these questions are rarely achievable through a single study design. Broadly, the analysis of how people use a tool for visual data analysis falls under the VDAR evaluation scenario described by Lam, et al.~\cite{lam_empirical_2012} and uses methodologies such as case studies for insight-based evaluations. Measuring user perception, conversely, targets methods for evaluating user experience through usability testing and questionnaires, with a focus on novice users. Finally, any measurement of user performance requires comparative tools for benchmarking. Each of these methods reveals rich and varied insights on the design of the system.  

In this article, we describe an evaluation that uses a combination of the above methods to understand the strengths and weaknesses of Tangere, but it does so in the context of initial usage and discoverability. The metrics we utilize in the study include discoverability, initial utility, and perceived ease of use and learning. 

\section{The case for no training in evaluations}
The advent of web and mobile-based software systems in recent years has considerably evolved our software development practices. We have witnessed a large-scale adoption of distributed development practices and platform-agnostic design approaches, and a significant shift in the software distribution and deployment mechanism. In the era of PCs, access to specialized software was limited to a niche audience, both due to the high cost of adoption and the expertise needed to use the systems. The market consisted of developers selling packaged software. Developers were able to curate the installation and usage experience, and spent a significant effort crafting peripheral resources, such as how-to guides and manuals. Often, users were exposed to this content during the software installation process.

This method of software delivery has largely been replaced by centralized repositories and distribution channels called app stores~\cite{basole_value_2012,noauthor_5_nodate}. Although the format of installation is still on-demand, there are no barriers to entry. Software applications are available to every one, and in most cases, available for free at the start. Although this distribution mechanism became mainstream with mobile operating systems, the adoption has also proliferated to desktop-based software systems.

In the app store model, 
the reduced barriers to entry alter the demographic of the target users. A potential user could have considerably less expertise and the ability to access the application at low to zero initial cost. This leads to the situation where users try multiple \textit{free} softwares that have the potential to fit their use-case before selecting one. Consequently, people have less patience for new software that may seem unintuitive at first use. Concurrently, app developers possess much less control on the installation experience (app installs on handhelds are instant) and can only build around the accepted practice of not overwhelming users with tutorial screens at first launch.

In this new framework of software development and distribution, we argue that capturing the behavior of untrained users becomes a key component of assessing likelihood of adoption. Two instruments available to capture this behavior are discoverability and initial utility.
 
We lean on this argument to propose altering the evaluation methodologies employed for visualization systems. Decades of research in the visualization field resulted in techniques and systems that were built within the framework of packaged software. Thus, the rationale to employ training as a part of evaluation was valid. As we usher into the new age of software distribution and consumption, it seems paramount to adopt approaches that are better tuned to evaluate initial behavior with software.


\section{Discoverability-based User Study}
As discussed above, we broadly sought to understand how people perceived the design of Tangere's interface and the interactions it employs. To benchmark the system, we compare people's performance with Tangere to that with an existing system --- Tableau's Vizable. Vizable is a publicly available system for analyzing data using linecharts and barcharts on a tablet device. While other tablet-based tools~\cite{noauthor_roambi_2014,noauthor_business_nodate, noauthor_jmp_nodate,jeffrey_m._rzeszotarski_kinetica:_2014} have also been presented in research and commercially, and were candidates to study, we chose Vizable because it most extensively supports the range of usage scenarios that Tangere does and it was readily accessible for use in the study. Furthermore, Tableau has been a leader in the consumer data-visualization application space. We, thus, felt that Vizable provided a solid standard to benchmark against.


The interface and interaction design of the two systems are considerably different, however. 
To design an experiment to robustly compare the two systems, we first identified the differences and similarities between them. These include:
\begin{enumerate}[\hspace{0em}1.]
	\itemsep-0.1em
	\item Unfamiliar users: Tangere was designed with unfamiliar users in mind, that is, people who have not used data visualization systems on tablets before~\cite{sadana_designing_2014}. Vizable, clearly, targets a similar audience~\cite{noauthor_vizable_2015}. Consequently, being learnable and discoverable is an important design goal for both. 
	\item Visualizations and views: Both systems visualize tabular datasets and require cleanly formatted data. They provide barchart and linechart visualizations, but Vizable omits a scatterplot. Vizable allows only one visualization on the screen at a time, while Tangere supports multiple-coordinated views.
	\item Task support: Both systems support a wide range of visualization tasks, such as filtering, sorting, and zooming, and omit others, such as methods for editing data. Task support also differs between the two tools. An example is selection --- Vizable does not support glyph selection whereas Tangere provides several methods for selecting glyphs. 
	\item Feature richness: Vizable supports certain features that are vital for commercial applications, but are currently missing from Tangere. Examples include operations such as undo and search, and features such as visualization sharing and bookmarking. 
	\item Interface and interaction design: The two systems also differ in terms of interface design. Vizable presents views in tabs, while Tangere presents them in a multiple-coordinated view configuration. The interactions employed for performing basic operations are also different across the two systems. Further, in Tangere, the emphasis is on presenting an overview. Thus, in a barchart with a very large number of categorical attributes, all bars are visible but appear extremely thin. Consequently, users have to zoom in to select an individual glyph, which adds steps. Conversely, in Vizable, bars are always of the same thickness and, thus, may be placed outside the bounds of the view. To access the bars, the user may need to perform significant scrolling. Therefore, ensuring that each glyph is accessible is prioritized over providing an overview.
	
	
	\end{enumerate}
	
We sought to understand how these differences affect people's use and experiences. Specifically, the questions we aimed to address were: 
\begin{itemize}
	\itemsep-0.1em
	\item Which system contains features that are more discoverable?
	\item Are people more accurate and productive when performing different types of tasks on either system?
	\item What are people's perceptions of the two systems? Does one feel more difficult to learn and use? 
\end{itemize}
To address the above questions, we identified four relevant metrics for comparison --- discoverability, initial utility, ease of learning, and ease of use.

\begin{table}[]
\centering
\begin{tabular}{|p{8cm}|}
\hline \\
\textbullet ~Which product was sold in the transaction that generated maximum profit?                                                                             \\
\textbullet ~On an average, how much profit was earned from each transaction of the products of type `herbal tea'?                                                \\
\textbullet ~How many products are of type coffee? What are their names?                                                                                          \\
\textbullet ~Which product types have sold the most and the least number of times? How about in the month of August?                                              \\
\textbullet ~Is there a relation between the profit and sales?                                                                                                    \\
\textbullet ~On which date did California's highest sales transaction occur?                                                                                      \\
\textbullet ~For the year 2014, what were the total number of transactions made?                                                                                  \\
\textbullet ~Which specific product type has most number of transactions with sales of 500\$ or more?                                                             \\
\textbullet ~How have the sales for the different regions changed over time?                                                                                      \\
\textbullet ~Which year was the most profitable? Which month within that year in particular? Does the same month tend to be the most profitable across all years? \\
\textbullet ~Is there a yearly pattern to the profits?                                                                                                            \\
\textbullet ~Is there a product type whose sales have been increasing over time? \\                                                                                 \\ \hline
\end{tabular}
\caption{Tasks given to participants for the beverage-sales dataset.}
\vspace{-1.5em}
\label{tab:tasks}
\end{table}

\subsection{Methodology}
This study followed a within-subjects design. Participants used both systems, exploring and completing tasks on one before using the other. To help us capture the discoverability of the systems' features, participants completed the tasks without receiving any training or tutorial on either interface. We modeled the tasks in the study on ones used by Drucker et al.~\cite{drucker_touchviz:_2013}, adopting the same two publicly available datasets (beverage sales \& superstore sales) that they used.\footnote{For a detailed description of the datasets, tasks and analysis, please refer to the supplemental material.} We employed a latin-square design to counterbalance order and dataset assignment: half the subjects experienced the beverage-sales dataset with Tangere and half experienced it with Vizable; half the subjects used Tangere first, while half the subjects used Vizable first. We video-recorded participants' interactions with the systems during the entire study. 

\paragraph{\textbf{Discoverability}} We captured discoverability by observing people use the systems for the first time. Participants were exposed to the systems without any training and given ten minutes at the start to explore the interface on their own. The systems were given to them in the default state that each opens in, with a placeholder dataset loaded.  
During this period, we monitored their interactions to record the features they discovered, but did so without participating in the exploration in any direct way.  
For this study, discoverability was less useful as a metric for directly comparing the features within the two systems since the systems have considerably different designs. Instead, we hoped to identify patterns for why operations were easy or difficult to discover and the implications of this within the corresponding system. 

\paragraph{\textbf{Initial utility}} After the initial exploration, participants received a series of tasks to complete with the system. They were given a total of 12 tasks and had 20 minutes to work. We apply the term ``initial utility'' as a metric here that measures the number of correct task completions in the context of the limited exposure to the system, as described above. 

To promote participants encountering all features in the system in the limited time, we paired several features together within a single task, e.g., selecting glyphs within a range of values (axis-based selection) was paired with filtering glyphs using `\textit{keep-only}' command. For the two datasets that we used, we designed matching sets of tasks, and presented those in a randomized order. The tasks we used for the beverage-sales dataset are presented in Table~\ref{tab:tasks}.

\begin{table}[t!]
\renewcommand\arraystretch{1.15}
\centering
\resizebox{\linewidth}{!}{
\begin{tabular}{|l|c|c|c|}
\hline
\textbf{Operation} & \textbf{\begin{tabular}[c]{@{}c@{}}Visual \\ Cue\end{tabular}} & \textbf{\begin{tabular}[c]{@{}c@{}}Interaction \\ Type\end{tabular}}  & \begin{tabular}[c]{@{}c@{}}\textbf{Count} \\ (OE/TC)\end{tabular} \\ \hline
\textbf{High} & \multicolumn{1}{l|}{} & \multicolumn{1}{l|}{} & \multicolumn{1}{l|}{} \\
Add charts & y & basic & \textbf{16} (16/0) \\
Change attribute & y & basic & \textbf{16} (16/0) \\
Filter through menu & y & basic & \textbf{16} (16/0) \\
Access data table & y & basic & \textbf{15} (15/0) \\
Change aggregation & y & basic & \textbf{15} (9/6) \\
Select by lasso & n & basic & \textbf{13} (11/2) \\
Select by tap & n & basic & \textbf{13} (9/4) \\
Toggle filter badge & n & basic & \textbf{13} (9/4) \\
Brushing and Linking & -- & -- & \textbf{12} (10/2) \\
Filter through selection & y & basic & \textbf{12} (7/5) \\ \hline
\textbf{Medium} &  &  &  \\
Remove a chart & n & basic & \textbf{11} (6/5) \\
Reset all filters & y & basic & \textbf{10} (7/3) \\
Remove filter badge & n & basic & \textbf{10} (2/8) \\
Select by rectangular window & n & basic & \textbf{9} (6/3) \\ \hline
\textbf{Low} &  &  &  \\
Zoom & n & basic & \textbf{6} (4/2) \\
Data count & y & -- & \textbf{5} (4/1) \\
Split line chart by category & y & basic & \textbf{5} (1/4) \\
Zoom by axis & n & basic & \textbf{4} (4/0) \\
Pan chart & n & compound & \textbf{1} (1/0) \\
Sort bar chart & n & compound & \textbf{1} (1/0) \\
Preview X or Y axis graph & y & basic & \textbf{0} \\ \hline
\end{tabular}}
\caption{Discoverability score for key features in Tangere. Visual Cue indicates whether the interface contains a visual affordance for the feature. Interaction type lists whether the interaction requires a single action (e.g. tap, pan, pinch) or combination of actions (e.g. tap-and-drag). Count represents the number of participants (out of 16) who discovered the feature: OE - discovered during open-ended exploration, TC - discovered during task completion.}
\label{tab:study1disctangere}
\vspace{-1.5em}
\end{table}

Participants were encouraged to think-aloud. This provided insight into how they were interpreting the applications. It was particularly useful for highlighting the interaction scenarios when their actions on the screen did not cause the system to respond in a manner they expected or when they correctly identified the operation they needed to use, but could not locate it on the screen. 

\paragraph{\textbf{Ease of learning and Ease of use}} We also captured participants' subjective feedback on the usability of the system through a survey at the end of the session. The survey consisted of 22 likert-scale questions largely based on the questionnaire generated by Elliott~\cite{elliott_grounded_2002}. For each system, participants completed the survey immediately after they finished the tasks with the system. 

\begin{table}[t!]
\renewcommand\arraystretch{1.25}
\centering
\resizebox{\linewidth}{!}{
\begin{tabular}{|l|c|c|c|}
\hline
\textbf{Operation} & \textbf{\begin{tabular}[c]{@{}c@{}}Visual \\ Cue\end{tabular}} & \textbf{\begin{tabular}[c]{@{}c@{}}Interaction \\ Type\end{tabular}}  & \begin{tabular}[c]{@{}c@{}}\textbf{Count} \\ (OE/TC)\end{tabular} \\ \hline
\textbf{High} &  &  & \textbf{} \\
Access both charts & y & basic & \textbf{16} (16/0) \\
Open filter menu & y & basic & \textbf{14} (14/0) \\
Change attribute from menu & y & basic & \textbf{14} (11/3) \\
Change aggregation from menu & y & basic & \textbf{14} (6/8) \\ \hline
\textbf{Medium} &  &  & \textbf{} \\
Select on Linechart & n & basic & \textbf{11} (8/3) \\
Filter from menu & y & basic & \textbf{11} (7/4) \\
Undo & y & basic & \textbf{10} (9/1) \\
Filter with swipe & n & basic & \textbf{10} (8/2) \\
Zoom Linechart & n & basic & \textbf{10} (5/5) \\
Sort bars & n & compound & \textbf{7} (6/1) \\
Toggle filter badge & n & basic & \textbf{7} (4/3) \\ \hline
\textbf{Low} &  & & \textbf{} \\
Remove filter badge & n & basic & \textbf{6} (4/2) \\
Change attribute with swipe & n & basic & \textbf{5} (3/2) \\
Add column with menu & y & basic & \textbf{5} (0/5) \\
Add column with pinch & n & basic & \textbf{4} (2/2) \\
Change aggregation with gesture & n & basic & \textbf{1} (0/1) \\
Switch attribute type (Q $\leftrightarrow$ C) & n & compound & \textbf{0} \\ \hline
\end{tabular}}
\caption{Discoverability score for key features in Vizable. Columns same as Table~\ref{tab:study1disctangere}.}
\label{tab:study1discvizable}
\vspace{-1.0em}
\end{table}

\subsection{Participants}
We recruited 16 participants (10 male, 6 female) for this study. The study took one hour to complete and participants were compensated with a \$20 Amazon Gift Card. The study took place in a usability lab. Participants performed the tasks using instances of the systems running on an Apple iPad Air 2 (the iPad did not have a case or stand). They could either hold the tablet in their hands or place it on a table. 

Participants were graduate students in Computer Science at our school. We sought to recruit participants who were \textit{subsequent learners}~\cite{davis_effect_1998}, i.e., people who are novices to a specific software system, but experienced with similar systems. This ensures that they not only have the required domain knowledge, but also a general understanding of which tools and functions will be available. This also guarantees that their performance is primarily a measure of the quality of the interaction and not of their subjective lack of understanding of the underlying features.

In particular, we sought experience with data visualization systems such as Tableau or Spotfire. We recruited participants through the roster of past courses on Information Visualization taught at our school to achieve that goal. 
None had previous experience with touch-based visualization systems, however.

\subsection{Results}
\paragraph{\textbf{Discoverability}} Tables~\ref{tab:study1disctangere} \&~\ref{tab:study1discvizable} present the discoverability results for a subset of features for the two systems.\footnote{The scores for all features of both systems are included in the supplemental materials.} Each table lists a count of the number of participants who discovered the different operations provided by a system. For Tangere, a majority of features achieved medium or high discoverability (7 or more participants discovered the feature). For Vizable, the results were similar with most of the key features achieving medium or high discoverability. 

\paragraph{WIMP vs. gesture} To examine the attributes of the design that assisted or hampered discoverability for novices, we first employed a division of WIMP-style versus gesture-based operations. This division helped explain a significant portion of behavior we observed. Unsurprisingly, the primary factor to positively affect discoverability was simplicity of action. Operations that used basic interactions (tap, pan, and pinch) or had clear visual cues were easily discoverable. Conversely, features that were invisible (e.g. filter menu), those that used complex interactions (e.g. hold+drag), or ones that required contextual operation (e.g. drag directly on axis) were difficult to discover. 

If designing for unfamiliar users, the results seem fairly intuitive. However, these results conflict with those observed in another study~\cite{drucker_touchviz:_2013}, where trained participants were both better at and preferred gesture-based operations over WIMP-based operations. As that study's authors point out, efficiency with gestures may not translate into discoverability and learnability.

\paragraph{Contextual gestures} Features that required contextual actions did not perform well on discoverability. A common example for both system includes contextualized drag gestures. These gestures require touch to occur at specific locations, followed by movement in specific directions. In Vizable, people can switch the attribute in a barchart by swiping left on the column header or change the aggregation by swiping down on it. In Tangere, people can drag a finger on the axis to create a range selection. Both these features had low discoverability, and were often discovered only serendipitously. The corresponding visual elements in both systems do not contain affordances that indicate the presence of these features. 

The design of such affordances is a non-trivial task, however.  Further, as we observed, once discovered, the operations are easy to learn and operate, further reducing the incentive for more explicit affordance. The solution Vizable adopts is to add redundancy --- besides using the swipe gesture, the attribute and the aggregation can also be changed from within a menu. Performing this action takes more steps, but certainly supports the use case of unfamiliar users. 

Although redundancy is an option, the feature density in visualization systems makes its pervasive application difficult to achieve without adding complexity and potentially compromising usability. Thus, we speculate that a better solution would be to redesign the feature to improve both the affordance and the feedback for the users.

\begin{figure*}[t!]
	\centering
	\includegraphics[width=1.5\columnwidth]{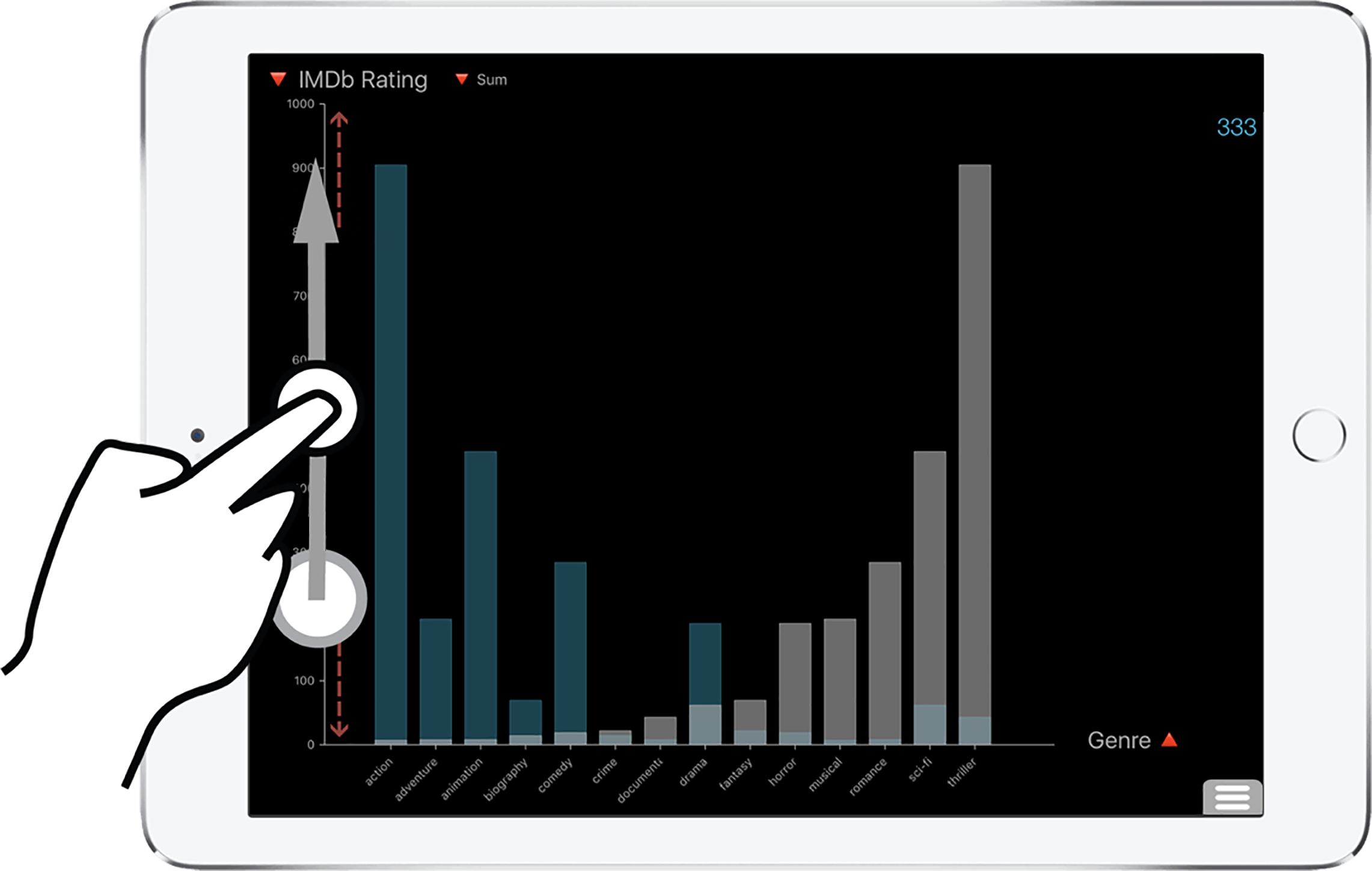}
	\caption{The barchart sort gesture in Tangere uses a \textit{hold + swipe} gesture, where the user has to hold a finger on an axis for 500 ms, and then swipe along the axis for ascending or descending sort. Feedback is provided using gray, translucent preview bars that animate to the final positions.}
	\label{fig:TangereSort}
  \vspace{-1em}
\end{figure*}

\paragraph{Complex actions} In Tangere, only a few participants could discover sorting on barchart (Figure ~\ref{fig:TangereSort}). The gesture used for sorting was hold-and-swipe on the axis~\cite{sadana_designing_2016}, and participants likely did not discover it because of its uniqueness within Tangere and to iOS in general (apps such as Google Maps and Apple Maps use it for zooming, but the usage is redundant because zooming can also be performed with the pinch gesture). Instead, one participant tried to sort the bars by sorting the data table, while two participants attempted to manually drag the tallest bars to one end (i.e. by demonstration~\cite{saket_visualization_2017}). 



\paragraph{Exploration approach} We observed participants take two broad approaches in exploring the system. The first set of participants spent a majority of their time interacting with features that they anticipated would exist, and found working the way they expected. Such features included lasso-based selection and filters. The second set of participants were more experimental. They quickly moved beyond what they already knew and tried to discover the other features of the systems. While the second set ultimately discovered more of the features, observing participants in the first set provided us with better insights into what people expect in such a system and how they translate their existing knowledge to new environments. 



\begin{table}[t!]
\renewcommand\arraystretch{1.5}
\centering 

\resizebox{\linewidth}{!}{
\begin{tabular}{l|c|c|c|}
\cline{2-4}
 & \multicolumn{3}{c|}{Average Score} \\ \hline
\multicolumn{1}{|l|}{System} & \textbf{Overall} & 1.Tangere 2.Vizable & 1.Vizable 2.Tangere \\ \hline
\multicolumn{1}{|l|}{Tangere} & \textbf{6.7} & 8 & 5.6 \\ 
\multicolumn{1}{|l|}{Vizable} & \textbf{5.4} & 6.4 & 4.4 \\ \hline
\end{tabular}}
\caption{Study 1: The average score of the participants (out of 12) for the two systems. The final two columns present scores adjusted for order effects.}
\label{tab:study2accuracy}
\vspace{-1.5em}
\end{table}

\paragraph{\textbf{Initial Utility}}  As discussed earlier, we use the metric \textit{initial utility} to refer to task completion success in the context of discoverability, i.e., a lack of formal training. To measure this attribute, we used a scoring system where participants earned one point for each task completed correctly, a half point for tasks attempted but completed incorrectly, and zero for a task not attempted. Table~\ref{tab:study2accuracy} presents the average scores participants achieved on the two interfaces, accounting for the order of use. Results indicate that participants scored slightly higher on Tangere than on Vizable (mean score 6.7 and 5.4, respectively). 
The order in which the systems were used did not affect the performance result ($F(1,28)=11.31, p<0.01$). Performance was better on Tangere, irrespective of whether it was used first or second. However, scores were significantly higher for participants who used Tangere first ($t(7)=-3.67, p<0.01$). We attribute the divergence in scores to the difference in discoverability of the two systems. As Tables~\ref{tab:study1disctangere} \&~\ref{tab:study1discvizable} present, compared to Vizable, Tangere had a larger number of critical features that were highly discovered.

For the above scoring calculation, we assigned each task the same value irrespective of its difficulty. This has the benefit of simplifying the comparison. However, since participants' performance was more nuanced and directly impacted by the difficulty of each question, a comparison of their performance for each individual task is more illustrative and clear. Here, the difference in scores is notable for two classes of questions.

First, questions that required quantitative filters received better scores in Tangere (e.g., \textit{Which product has most transactions with sales of \$500 or more?}). This was mainly because Vizable did not provide an easy method to filter on quantitative attributes. In Vizable, the dynamic query filters could only be applied to categorical attributes. To filter on numerical values, users had to transform a quantitative attribute into a categorical one and, in absence of multiselection on bars, individually filter out the bars for each value. In Tangere, conversely, filters could be applied to quantitative attributes both using selection + `keep only/remove' and dynamic query widgets.

Second, on questions that required analyzing patterns over time, participants fared better on Vizable than on Tangere (e.g., \textit{Is there a product type whose sales have been increasing over time?}).  On both Vizable and Tangere, one could best answer these questions using a linechart. However, participants found the implementation of a linechart in Vizable to be better and considerably more usable than in Tangere. In Vizable, the linechart used semantic zooming, thus the view at each state displayed aggregated values. In Tangere, the linechart displayed each item individually, aggregating only at the level of each day (Figure~\ref{fig:interface}). This view was noisy, and thus, less useful for participants. Participants found Vizable's implementation to be more approachable and more usable.

\begin{table}[t!]
\renewcommand\arraystretch{1.5}
\centering
\resizebox{\linewidth}{!}{
\begin{tabular}{|l|l|c|c|c|c|c|c|}
\hline
\multicolumn{2}{|l|}{\multirow{2}{*}{\textbf{System}}} & \multicolumn{3}{c|}{\textbf{Ease of Learning}} & \multicolumn{3}{c|}{\textbf{Ease of Use}} \\ \cline{3-8} 
\multicolumn{2}{|l|}{} & $M$ & $\bar{x}$ & $\textit{sd}$ & $M$ & $\bar{x}$ & $\textit{sd}$ \\ \hline
\multirow{2}{*}{Overall} & Tangere & 4 & 3.8 & 0.6 & 4 & 3.6 & 0.7 \\
 & Vizable & 3 & 3.0 & 0.7 & 3 & 2.8 & 0.8 \\ \hhline{|=|=|=|=|=|=|=|=|}
\multirow{2}{*}{Vizable First} & Tangere & 4 & 3.5 & 0.6 & 4 & 3.4 & 0.7 \\
 & Vizable & 3 & 2.9 & 0.6 & 3.5 & 2.9 & 0.8 \\ \hline 
\multirow{2}{*}{Tangere First} & Tangere & 4 & 4.1 & 0.5 & 4 & 3.9 & 0.5 \\
 & Vizable & 3 & 3.0 & 0.8 & 3 & 2.7 & 0.9 \\ \hline 
\end{tabular}}
\caption{Study 1: Participants' response to the ease of use and ease of learning questionnaire for the two interfaces. The table presents overall scores and scores adjusted for  order effects. Scale used: 1: Strongly Disagree, 5: Strongly Agree}
\label{tab:study2eoleou2}
  \vspace{-1.5em}
\end{table}

\paragraph{\textbf{Ease of learning and Ease of Use}}
Table~\ref{tab:study2eoleou2} 
summarizes participants' ratings for the  systems on two subsets of questions that captured Ease of Learning and the Ease of Use opinions. Overall, participants rated Tangere higher on ease of learning than Vizable, regardless of the order. The difference in scores between the systems is significant (Wilcoxon's test: $Z=2.75, p<0.01$). Similarly, Tangere's rating for ease of use was also significantly higher than Vizable's. (Wilcoxon's test: $Z=2.83, p<0.01$), with the difference being particular notable for participants who used Tangere first. 

We attribute the disparity in scores to two factors --- the difference in how the systems respond to elementary interactions, and the difference in how they support low-level tasks. In Vizable, not every element on the screen responds to actions such as tap and swipe, which tend to be the predominant actions participants perform when experimenting with an interface. Conversely, in Tangere, interactions such as tap, double-tap and swipe are operable on most of the elements. As a result, the interface feels more responsive and accessible.

Similarly, differences also exist in how the tools support low-level tasks. As discussed earlier, Vizable does not support selection of glyphs in a barchart. Thus, the only interaction possible on the bars is swipe. Consequently, most of participants' actions were performed on the filter menu on the left, making filters seem more relevant to the workflow than the tasks demanded. In other words, the interaction design assigned an importance to the features that may not match the actual order of relevance of the features. This may have caused participants to perceive the system to be more complicated to use. 

\begin{table}[ b!]
\centering
\renewcommand\arraystretch{1.25}
\begin{tabular}{|p{0.55\linewidth}|c|c|}
\hline
\textbf{Question} & \textbf{Tangere} & \textbf{Vizable} \\ \hline
Which system was easier to learn? & 14 & 2 \\
Which system was easier to use? & 13 & 3 \\
Which system was easier to \newline understand and interpret? & 11 & 5 \\
Which system has a more complex \newline interface ? & 6 & 12 \\ \hline
\end{tabular}
\caption{Study 1: Participants' preference for the two systems. The value in the cells represents the number of participants who answered Tangere or Vizable for the specific question.}
\label{tab:study2preference}
  \vspace{-1.5em}
\end{table}

At the end of the study, we also asked participants to directly compare the two systems. Table~\ref{tab:study2preference} presents their preferences between the systems on ease of learning, use, and complexity. Overall, participants preferred Tangere to Vizable, finding it to be both easier to learn ($\chi^2 = 7.56, p=0.006$) and easier to use ($\chi^2 = 5.06, p=0.02$). 
Participants primarily differentiated the system on intuitiveness and aesthetic appeal, as illustrated in the comments below. 

\begin{myquote} 
P13: ``\textit{I liked Tangere better because it was easier to use and I understood faster how to use it. Vizable was prettier but wasn't very useful. It was also less intuitive than Tangere.}" \\ 
P1: ``\textit{I felt like (Vizable) compromised aesthetics for function. Often times I wasn't sure what I could or should do next}." \\
P9: ``\textit{I liked the intuitive interface and flexibility of Tangere. I was able to quickly learn how to navigate the features, and was able to find a way to answer all the questions in the task.}"
\end{myquote}


\noindent
Beyond the participants' summary subjective opinions, we also observed a number of unique perceptions about each system throughout the sessions.

\noindent
\textbf{Vizable}

\begin{itemize}
\item Participants were confused between Vizable's filter and change attribute panels. They appeared at the same location with the same slide-in animation, used the same visual style, and contained the same list of attributes. Thus, it was easy to mistake one for the other. 
\item In the linechart, Vizable uses bars as glyphs for representation, with the line in the chart connecting the top of the bars. The presence of bars in the linechart was confusing for several participants as there already was a bar chart. 
\end{itemize}
		
\noindent\textbf{Tangere}

\begin{itemize}
\item Participants appreciated being able to brush between views, even though the questions were designed to not need the use of the feature, given it was missing in Vizable. 
\item The data-table, available in Tangere but missing in Vizable, was used extensively and participants universally appreciated being able to access it. 
\item A subset of the participants questioned the choice of colors in Tangere, expressing a preference for a white background, as Vizable provides, over a black one. 
\end{itemize}

 \begin{figure}[t]
 	\begin{subfigure}[b]{.95\linewidth}
 	\includegraphics[width=\linewidth]{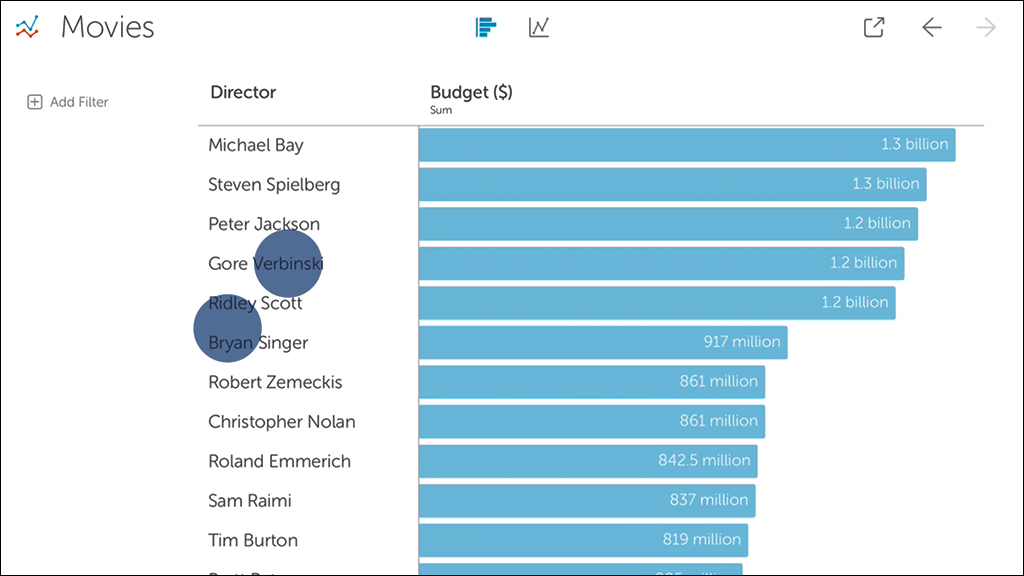}
 	\caption{Pinch gesture is initiated on a column.}
 	\label{fig:countlabel}
 	\end{subfigure}
 	\par\smallskip
 	\begin{subfigure}[b]{.95\linewidth}
 	\includegraphics[width=\linewidth]{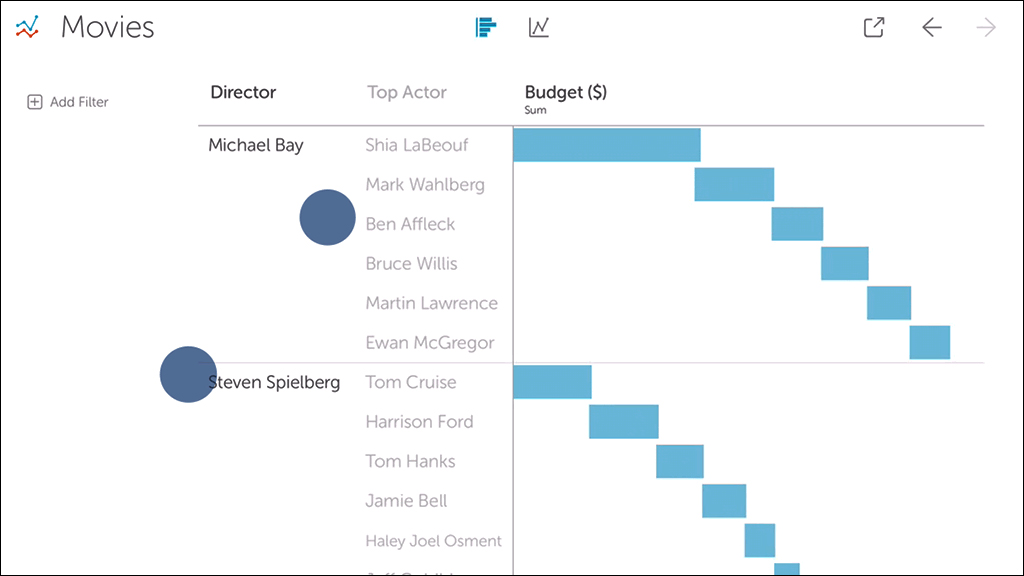}
 	\caption{Pinching out reveals a second column.}	
 	\label{fig:countlabelvizable}
 	\end{subfigure}
 	\par\smallskip
 	\begin{subfigure}[b]{.95\linewidth}
 	\includegraphics[width=\linewidth]{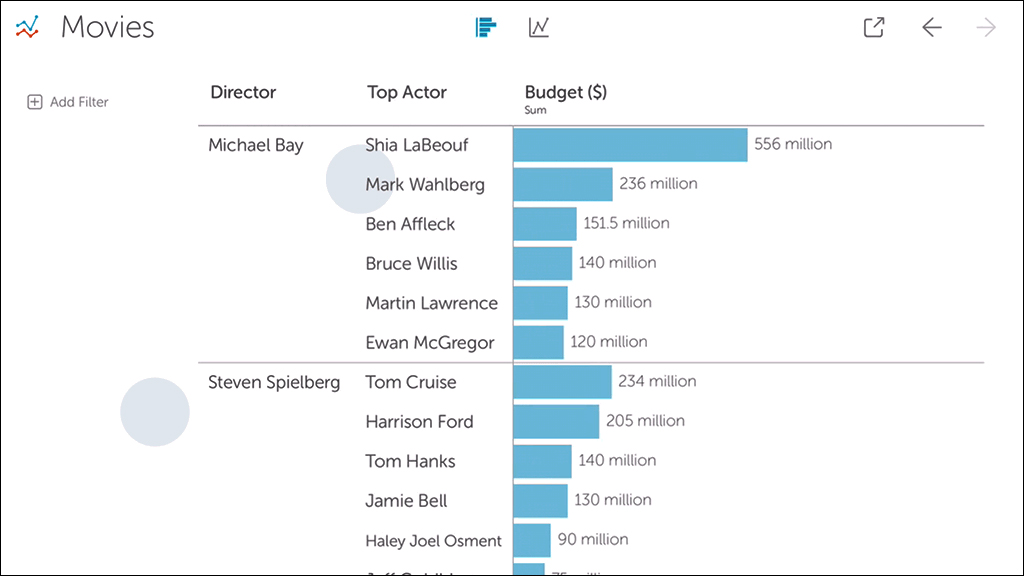}
 	\caption{Bars divide with the new attribute as a subcategory.}	
 	\label{fig:countlabelvizable}
 	\end{subfigure}
 	\caption{Vizable: Three stages of a pinch gesture to add a categorical attribute as a column. A quantitative attribute can similarly be added by pinching directly on the bars.}
 	\label{fig:VizablePinch}
   \vspace{-1.5em}
 \end{figure}

\subsection{Observations}
The above results present a fairly encouraging review of Tangere. In particular, several aspects of Tangere's design resonated more with participants than Vizable. The number of participants in the study was fairly small, however, and thus we feel it would be overreaching to take away too much from it in general. Instead, we sought to uncover the low-level, nuanced differences in the design of the two tools that influenced the participants' preferences. Below we describe such observations that emerged from the study.

\paragraph{\textbf{Hacking discoverability}} A key factor that affects discoverability is familiarity. Familiarity leads to people seeking an expected operation or attempting a known interaction in a new environment. We typically use this property to replicate common behavior. However, it can also be used as a ``hack'' to force a desired outcome. For instance, in Vizable, the designers opted to use a pinch-to-zoom gesture to add columns to a barchart (Figure~\ref{fig:VizablePinch}). Such a use of the gesture is fairly unconventional, since the degree of compatibility between the pinching action and add-column behavior is very low~\cite{beaudouin-lafon_instrumental_2000}. 
We argue that the resulting design suffers from low usability, as it lacks any affordance or indication that the pinch gesture would result in the addition or removal of a column of bars. Participants' confusion with the operation was apparent. P5 mentioned ``\textit{The zoom-in/zoom-out feature for barchart is really weird. I understood it is used to get more insight on sub-column (category $\rightarrow$ subcategory) but sometimes I got new unwanted columns''}. It is notable that the participant describes the operation as `zoom-in/zoom-out' instead of `pinch gesture', clarifying the reason for their confusion with the interaction.

Ultimately, Vizable attempts to leverage people's familiarity with the pinch interaction. By combining it with an outcome that is expectedly comprehensible, the goal is to result in a feature that can be serendipitously discovered. Given the frequency with which participants are likely to use the operation, the behavior can also be learned. 
However, it is difficult to argue for this style of design. The use for add-column is both different from the form of the actual gesture (the pinch action), and from the familiar and accepted use of the interaction.

\paragraph{\textbf{Consistency}}
Consistency is often cited as a theme essential to good design~\cite{norman_design_2002}. However, Tangere and Vizable adopt very different approaches to consistency. For Tangere, maintaining consistency between techniques was fundamental to the design process~\cite{sadana_designing_2016}. The three visualization techniques in the system support the same set of operations, and use the same interactions to invoke those operations throughout. Vizable's approach to consistency is considerably different. Its two visualizations, barchart and linechart, exist independently in separate tabs and do not share any common operations or interactions. While the linechart supports selection and zoom through tap \& drag and pinch, the barchart does not support either of these operations and instead provides filtering, sorting, and adding a column or bars using a similar combination of gestures. 

Naturally we were interested in assessing if the two approaches resulted in a different experience. Our analysis suggests that the lack of consistency within Vizable negatively affected participants' perception of the system. This is because spending time on one chart did not prime participants to use the other chart. Instead of translating the knowledge from the first chart, participants had to spend time and effort identifying and learning the operations in the second chart. It was also  one of the key reasons participants found the interface to be complex, non-intuitive, and difficult to learn (P14: ``\textit{Learning period in the Vizable system is a lot. Some of the features are not apparent immediately and intuitively}.").

Inconsistency also has other drawbacks besides complexity and reduced learnability. An inconsistent behavior introduces a perception of dissociation or disjointedness between the fragments of the system, wherein people might understand the views as separate subsystems. For instance, participants were unclear if operations such as selection or filtering translated across the two views in Vizable. The same confusion was not apparent in participants' use of Tangere. Issues such as these are significant since they
 would only be amplified when the system is expanded to more techniques, or views that are colocated and coordinated, such as in Tangere. 

\begin{table}[b!]
\renewcommand\arraystretch{1.15}
\centering
\begin{tabular}{lcc|}
\cline{2-3}
\multicolumn{1}{l|}{} & \multicolumn{2}{c|}{Order} \\ \cline{2-3} 
\multicolumn{1}{l|}{} & \multicolumn{1}{c|}{First} & Second \\ \hline
\multicolumn{3}{|l|}{\multirow{2}{*}{\textbf{Vizable}}} \\
\multicolumn{3}{|l|}{} \\ \hline
\multicolumn{1}{|l|}{Filter with swipe} & \multicolumn{1}{c|}{3} & 7 \\ 
\multicolumn{1}{|l|}{Remove filter badge} & \multicolumn{1}{c|}{1} & 5 \\ 
\multicolumn{1}{|l|}{Toggle filter badge} & \multicolumn{1}{c|}{2} & 5 \\ 
\multicolumn{1}{|l|}{Change attribute with swipe} & \multicolumn{1}{c|}{1} & 4 \\ \hline
\multicolumn{3}{|l|}{\multirow{2}{*}{\textbf{Tangere}}} \\
\multicolumn{3}{|l|}{} \\ \hline
\multicolumn{1}{|l|}{Select by tap} & \multicolumn{1}{c|}{8} & 5 \\ 
\multicolumn{1}{|l|}{Select by lasso} & \multicolumn{1}{c|}{8} & 5 \\  
\multicolumn{1}{|l|}{Interpret data count} & \multicolumn{1}{c|}{5} & 0 \\ \hline
\end{tabular}
\caption{Study 1: Order effects in the discoverability metric. The value represents the number of participants (out of 8) who discovered the feature when the system was used first or second.}
\label{tab:study1ordereffects}
\vspace{-1.5em}
\end{table}

\paragraph{\textbf{Order effects}} In capturing discoverability, we observed interesting order effects. Table~\ref{tab:study1ordereffects} presents features that were discovered by at least three ($\sim$20\%) additional participants in one order condition than in the other. Results indicate that the difference occurred exclusively in the condition when Tangere was used first. In other words, using Tangere first \textit{increased} the number of features participants discovered in Vizable. Conversely, using Vizable first \textit{reduced} the number of features a participant discovered in Tangere.

The difference, in all likelihood, emerges from a change in user expectations. Using Tangere first primed participants to expect certain features and interactions within Vizable. For instance, the filter badges in Tangere were interactive and could be tapped and  horizontally swiped to toggle and remove the filter, respectively. Consequently, more participants attempted these interactions on the filter badges in Vizable when the tool was used second compared to when it was used first. In general, Tangere more broadly supports interactions directly on glyphs. We speculate that this led to participants more easily discovering features such as filtering and changing attributes by swiping on interface elements.
Using Vizable first similarly affected scores for Tangere. Since Vizable does not support selection, fewer participants attempted to select glyphs within Tangere, negatively affecting its discoverability. 

These observations are interesting as they explain the role of priming in improving discoverability. While we can leverage people's familiarity with existing interfaces, it would be detrimental to assume it. The order-of-use effects on discoverability clearly underscore the need for the features to be better designed. Through our focus on evaluating untrained users, we have been able to more critically evaluate and unearth deeper flaws in the design of Tangere.

\section{Discussion}

\subsection{Implications for evaluation design}
The big differentiator in our approach to evaluation, compared to more traditional methods, was the lack of interface training in the study. 
In reflecting on the outcomes of the study, we believe that the discoverability and initial utility metrics resulted in observations that were richer and more actionable for the design of the system than traditional metrics. For instance, the key design takeaways of \textit{contextual gestures}, \textit{consistency} and \textit{gesture uniqueness} (presented later) are observations that could only emerge from scenarios where no training was provided. Given training, even if participants fail to use some operations, observations of potential design problems could be conflated by the potential insufficiency of training provided. In our method, no such confusion existed. 

We also considered other approaches to training, such as providing training after participants completed the self-discovery session but before they completed the tasks. However, in the five minutes of open-ended discovery, participants often discovered only a subset of features available in the system. As Table~\ref{tab:study1disctangere} highlights, a significant number of features were discovered during task completion. Providing training would have prevented us from capturing this information.

\subsubsection{Evidence for discoverability}
A challenge we faced in measuring discoverability was agreeing on cues to use to mark a feature as ``discovered.'' Often, features were discovered serendipitously either through accidental or imprecise touches on the screen. For example, one could drag within a chart to activate lasso selection, drag on the axis to enable axis-based selection, and drag on the edges to bring in side-view panels. Since the activation regions were connected, participants would often accidentally activate one operation when attempting another. 

In such situations, it was unclear if we should mark the accidentally activated feature as \textit{discovered}. We initially considered requiring an acknowledgement from the participants, such as them reattempting the feature or through verbal confirmation that they recognize the feature. This appeared to be only semi-effective, however. It was useful for unfamiliar features, such as axis-based selection or sorting, for which participants would often reattempt the operation. For familiar features such as tap-to-select or lasso-select, however, participants would not necessarily reattempt the action, but it was difficult to ascertain if that was due to familiarity with the operation or lack of acknowledgment of it.
Ultimately, we marked a feature as discovered either when participants consciously acknowledged it after the first use, or used it again over the course of the exploration.

\subsubsection{Discoverability for inclusivity}
One potential assessment of the design of Tangere and the outcome of this evaluation is that the results seem fairly obvious. Specifically, one could argue that the outcome should have been expected because the differences between the tools are rather glaring. However, such an assessment has two drawbacks. 

First, it likely contains hindsight-bias. When we initiated this research, we did not anticipate the results to be skewed to such a large degree. Vizable represented the state-of-the-art in terms of publicly-available, scalable, tablet-based visualization systems.

Second, and more importantly, such an assessment undermines the significance of the results. Tangere and Vizable are systems that were carefully crafted to support a narrow set of features. It is possible that a study with initial training would produce results showing comparable performance by the two systems. In fact, in our own usage of Vizable, after the initial learning curve, we found the tool to be performant, consistent and fun to operate -- a testament to the effort that went into building it. 
However, this observation, which might also emerge from an evaluation with training, ignores the experience of a critical set of users---non-experts who receive no training. These people will have access to the software, but might find themselves unable to operate it. The set also includes people who use the system only intermittently, after significant gaps of time. With our approach using a discovery-based evaluation, we were able to be more inclusive. Our results highlight performance shortcomings of one design that have the potential to hinder the experience of a large number of people.

\subsection{Implications for future interface design}
Below we summarize key findings from the study as take-aways to assist the design of future systems in this area. We intend for these observations to complement the guidelines from HCI literature on designing systems for touch devices.
\begin{itemize}
  \setlength{\itemsep}{0pt}
  \item Contextual Gestures: Gestures requiring precise positioning and movement are difficult to discover. The effect is exacerbated in visualization interfaces since the density of elements that can be interacted with is fairly high. As much as possible, the use of such gestures should be limited or avoided altogether.
  \item Consistency: If the system presents multiple visualization techniques, whether concurrently or in separate spaces, employing the same interactions for the common operations considerably simplifies the learning required. Such a consistency of design has not been a major design goal of visualization applications for mobile and touch devices in the past, a key omission in our view.
  \item Repurposing gestures: Although the visualization domain is fairly unique compared to other domains, visualization applications need to align with the design practices employed elsewhere and must not alter the behavior of common gestures that users learn and use in other domains.
  \item Primary tasks and gestures: Even in a constrained environment of a tablet or a phone, it is important to maximize system support for the basic tasks, such as selection, and standard gestures, such as tap, pan, and pinch.
  \item Gesture Uniqueness: Non-standard, compound gestures (e.g. hold-and-drag), if employed uniquely, will suffer from low discoverability and memorability. New, innovative gestures may be desirable in a research contributions context, but we believe they have serious shortcomings from the point of view of pragmatic usage.
  \item Gesture Overloading: Gestures that are differentiated based on properties of movement (drag vs.~hold-and-drag, or swipe horizontally vs.~swipe vertically) are difficult to discover and confusing to remember. It is best to avoid employing such gestures.
\end{itemize}
We believe that developers of future data visualization applications for mobile devices will benefit from employing designs that follow these principles. No system in this space has achieved broad success yet. An inattention to designing for the these principles may, in part, explain the situation.

%

\section{Conclusion}
In this article, we presented a user study to evaluate Tangere, a visualization application for analysis of tabular data on a multi-touch tablet device. We captured initial usability of the features of the system by assessing usage of participants unfamiliar to the system. We also compared participants' performance to that using Vizable, a commercially available tablet visualization system. Results indicated that participants perceived Tangere more favorably than Vizable. 

Much further work remains. While our focus was primarily on gauging user perception towards the interactions within the tool and the tool at large, another useful approach would extract styles and patterns from people's usage of the tool for data analysis. Specifically, how do people construct insights on a touch-input based system and how does the tool assist in this process? 
Similarly, longer term deployment-study experiments would reveal richer insights on the limitations of the current design and the level of expertise achievable with the tool.

Our work takes an important first step in assessing the design and usability of two tablet-based visualization systems. We presented results from an approach that targets the experience of new users trying the system without any training. We hope that in the future, our results and reflections will help serve as guidelines to assist in the design of other, more diverse approaches towards exploring data on mobile and tablet devices.

\section*{Acknowledgements}
This work was supported by a Google Faculty Research Award and by the National Science Foundation via award IIS-1320537.


\bibliographystyle{abbrv-doi}

\bibliography{bibliography}
\end{document}